\documentclass{aa}  

\usepackage{graphicx}
\usepackage{txfonts}

\usepackage{placeins}           

\usepackage[final, defaultcolor = black, addedmarkup = bf, deletedmarkup = sout]{changes}

\begin{document}

   \title{Prospects for detecting surface color heterogeneity on asteroid surfaces from sparse multiband photometric survey data}

   \author{O. Humes\inst{1}
        \and J. Agarwal\inst{1}
        }

   \institute{Institut f\"{u}r Geophysik und Extraterrestrische Physik\\
             \email{oriel.humes@tu-braunschweig.de}}

   \date{Received February 12, 2026}

  \abstract 
  {Automated sky surveys frequently report sparse-in-time multiband photometric observations of asteroids passing through their fields of view. Photometric data are currently available for tens of thousands of asteroids, and new data collection is ongoing. }
   {We aim to describe and characterize the performance of a statistical test for identifying asteroids that display surface color heterogeneity based on sparse-in-time multiband photometric survey data.} 
   {Using simulated photometry for a set of synthetic asteroids with predetermined physical properties, we estimated the sensitivity of the statistical test for surface color heterogeneity to errors in assumed model properties using a Monte Carlo approach. We evaluated the detection and false positive rates as a function of the number of observations, measurement noise, error in assumed period, pole orientation, shape, and phase function. We examined the required accuracy in various parameters of the photometric model needed to obtain reliable results to evaluate the feasibility of applying the test to realistic datasets.}
   {Regional-scale surface color heterogeneity can be detected by examining differences in the shape of an asteroid's light curve as a function of viewing geometry across multiple bandpasses. Differences in light curve shapes as a function of wavelength are highlighted in this work through comparison of the observed photometric measurements to the predictions of a well-fitting, uniformly colored photometric model. Statistically significant deviations from the prediction of the uniformly colored model are taken as evidence of surface color heterogeneity. The performance of this test depends on the accuracy of model assumptions, with the detection rate being most sensitive to errors in the assumed rotational period, while the false positive rate is most sensitive to errors in the assumed band-dependent phase functions. }
   {Current photometric models derived from sparse photometry can achieve the required accuracy in model parameters for thousands of asteroids observed by present-day surveys.}

   \keywords{}
\titlerunning{Prospects for detecting asteroid surface heterogeneity}

   \maketitle
   \nolinenumbers

\section{Introduction}

In recent years, the volume of high-quality publicly available time series photometric data of astronomical objects has increased dramatically. This increase is in large part thanks to the efforts of ongoing automated astronomical surveys, including the Vera Rubin Observatory's Legacy Survey of Space and Time (LSST), the Zwicky Transient Factory (ZTF), and the Asteroid Last-Alert Terrestrial-impact System (ATLAS). These facilities produce multiband photometric measurements of large portions of the sky across multiple wavelength bands over long stretches of time while including moving Solar System objects in the foreground. The emerging field of time domain astronomy seeks to utilize these datasets to understand the evolution of astronomical phenomena over time. 

Time domain photometry of astronomical objects can be used to leverage the changing viewing and illumination conditions inherent to time series observations of moving rotating objects, such as asteroids and trans-Neptunian objects (TNOs), in order to infer the spatial distribution of color and albedo across an object's surface. Given knowledge of an object's orientation, temporal full-disk photometric coverage of unresolved point sources can be transformed into a probe of spatial variation. This principle has been used to determine the shapes and albedo distributions of asteroids from their light curves \citep{Kaasalainen-2001} and infer the distribution of spots on stellar surfaces \citep{Roettenbacher-2011, Roettenbacher-2017}, and it has also been proposed as a method to ``map'' the surfaces of exoplanets \citep{Kawahara-2010, Fujii-2012}. In the past, specialized time series observations of individual objects of interest enabled surface albedo, color, and compositional variations to be inferred for a small number of Solar System objects, including Iapetus \citep{Morrison-1975}, Pluto \citep{Drish-1995, Grundy-2001}, and Haumea \citep{Lacerda-2009}. Previous studies of spatial variability on small Solar System bodies from unresolved photometry have typically relied on observing campaigns dedicated to particular individual objects of interest. The present era of large-scale automated photometric surveys allows astronomers to ``cast a wider net'' and observe large numbers of asteroids over multiyear timescales. However, the data generated by automated surveys sample asteroid light curves at irregular, sparsely separated intervals in time, presenting novel challenges for their analysis and interpretation. Techniques, such as asteroid shape modeling from light curve inversion, have been successfully adapted to deal with the challenges of working with sparse datasets \citep{hanus-2011}. In this paper, we examine the feasibility of using multiband photometric data generated by large-scale surveys to detect asteroids with surface color variations, and we propose a statistical test to identify candidate asteroids displaying color heterogeneity and analyze its performance in the context of the expected performance of contemporary multiband photometric surveys (LSST, ZTF, ATLAS) and data analysis techniques. 

Detecting large-scale surface heterogeneity on asteroids allows these objects to be investigated as geological bodies. Several classes of asteroids and meteorites are thought to have been part of protoplanetary bodies that have undergone substantial internal differentiation via silicate melting, leading to large-scale compositional heterogeneity throughout the interior as a function of depth \citep{Yang-2007, Scott-2015-differentiated}. Modeling has also shown that early thermal metamorphism is able to produce a layered interior structure in icy bodies, with a relatively ice-rich shell surrounding an unmelted rockier core \citep{Castillo-Rogez-2010}. Subsequent collisions leading to the catastrophic breakup of layered protoplanets are expected to produce asteroid fragments exhibiting a diverse array of surface compositions, depending on which layer of the protoplanet each asteroid originated from \citep{Michel-2004}. These collisional products potentially include asteroids with mixed compositions, with their parent material originating from different compositional layers. Such asteroids, identifiable through their heterogeneous surfaces, would provide a relatively accessible example of a geophysically significant transition region (e.g., the core-mantle boundary) of a protoplanetary interior and inform our understanding of planetary formation. Such an explanation has been invoked to explain the surface heterogeneity on (16) Psyche, and it motivates the science questions of the NASA mission of the same name \citep{Elkins-Tanton-2020}. 

For asteroids that have undergone little to no internal differentiation (typically the ``primitive'' asteroids), we expect an overall homogeneous bulk composition, as they remain unaffected by global-scale geologic restructuring. However the near surfaces of these asteroids likely undergo color changes relative to their interiors as a result of their exposure to the space environment, either through the effects of radiation-induced space weathering \citep{Pieters-2016} or due to the formation of a volatile-poor rubble mantle \citep{Hsieh-2015}. For these asteroids, detection of a heterogeneous surface would indicate a recent exposure of subsurface material that has not yet equilibrated to the thermal and radiation environment of the near surface. Space weathering-related spectral differences controlled by variation in surface exposure age have been invoked to explain the observed spectral diversity of asteroid families \citep{Nesvorny-2005} and across surfaces of different ages on individual asteroids imaged by spacecraft \citep{DellaGiustina-2020S, Ishiguro-2007}. Quantifying the number of asteroids displaying surface heterogeneity in this case would provide an estimate for the frequency of subsurface-exposing events (such as impact cratering, collisional breakup, regional-level mass wasting, or rotational fission) and the timescale over which space weathering is able to modify the spectral characteristics of asteroid surfaces. Similarly, quantifying the range of colors present across individual bodies would provide insights into the magnitude of plausible color changes induced by space weathering and/or volatile loss among asteroids with the same overall bulk composition. In the case of volatile exposure, the detection of volatile-related surface heterogeneity informs our understanding of the formation and migration histories of asteroids with respect to volatile condensation lines. While unresolved multiband photometry is unlikely to be able to directly answer these science questions, the ability to efficiently identify asteroids exhibiting heterogeneous surfaces from a population of millions is needed in order to determine which asteroids represent the most promising targets for more detailed follow-up characterization (including through Earth-based observations and missions) that can uniquely enrich our knowledge of asteroids as geological bodies.

\section{Statistical test for detecting heterogeneity}

A fundamental challenge in deriving color information from sparsely sampled multiband photometry is the non-simultaneity of measurements across multiple photometric bands. One approach to solving this issue is to employ photometric modeling, whereby existing brightness measurements are fit with an idealized model of asteroid brightness that is then used to predict what the brightness of an object would have been during epochs in which it was not observed in the corresponding band. Such models include photometric properties (absolute magnitude and phase function) and physical properties (the shape of the asteroid and its rotational properties). Methods for fitting the free parameters of suitable photometric models of varying complexity to sparse photometric data can be found across the literature, as in, for instance, \citet{hanus-2011, Lu-2016, Cellino-2019, Carry-2024}. These existing techniques are able to produce suitable models when given sparse photometry alone. This approach effectively corrects for the changing brightness of an asteroid due to the changing viewing geometry and rotation, allowing its color to be estimated by subtracting the predicted brightness of the asteroid in each band under identical viewing geometry. Such a strategy has been used to derive mean color estimates from ATLAS \citep{Mahlke-2021}, ZTF, and LSST photometry \citep{Carry-2024}. 

We proceeded under the assumption that, for a given asteroid, we are able to obtain both a photometric model, $m_B(t)$, that predicts the magnitude of a given object in band $B$ and a rotational model (pole orientation and rotational period) that predicts the subsolar and sub-observer latitudes and longitudes in its corotating frame at any epoch $t$. An example of such a photometric model appears in Appendix A. The photometric model $m_B(t)$ should assume constant reflectance properties across the entire surface of the asteroid in each particular band, which implies that it should be spatially uniform in color.
The uniformly colored assumption acts as the null hypothesis--the following significance test evaluates how likely it is that an observed set of photometric data could arise from observations of an asteroid lacking surface variegation, with a sufficiently high significance indicating that the observations are unlikely to be attributable to a uniformly colored asteroid. In the model, the photometric properties (absolute magnitude, phase function) may vary from band to band, but the physical properties (the shape of the asteroid and its rotational properties) should be wavelength invariant.

To test for surface color variegation, we computed the difference between the observed magnitude of the asteroid and its predicted magnitude according to the uniformly colored model $m_B(t)$ for all observational epochs in their corresponding photometric band. The residual between the prediction and model is a function of time and bandpass $\Delta m_B(t)$. If an asteroid's surface varies in color, we expected that $\Delta m_B(t)$ will depend on the viewing geometry at the time of observation, depending on whether the inhomogeneity is illuminated and visible to the observer at that time. If the inhomogeneity is a consistent surface feature, the value of $\Delta m_B(t)$ is consistently offset from zero each time that part of the surface is illuminated and visible. To evaluate which side of the asteroid is illuminated and visible at any given epoch, we used the model's rotational spin axis, orientation, and rotational period to compute the unit vectors pointing in the direction of the Sun ($\hat{E_0}$) and the observer ($\hat{E}$) in the rotating coordinate frame of the asteroid (i.e., \citet{Durech-2010}). We could then consider $\Delta m_B(t)$ as a function of $\hat{E_0}$ and $\hat{E}$. We computed the average of the illumination and viewing angles $\hat{\bar{E}} = \frac{1}{2}\left(\hat{E_0} + \hat{E}\right)$ at each epoch and considered the residual function $\Delta m_B(\hat{\bar{E}})$. The direction $\hat{\bar{E}}$ approximately identifies, within the coordinate system of the asteroid, the area of the surface from which the greatest flux is observed given a particular illumination and viewing geometry. While the functions $\Delta m_B(\hat{E})$ and $\Delta m_B(\hat{E_0})$ can also be used, we found greater detection rates for spots using the average $\Delta m_B(\hat{\bar{E}})$, as a strongly illuminated facet does not contribute to the observed flux if it is not facing toward the observer, and vice versa. This consideration is particularly important for near-Earth asteroids (NEAs), for which $\hat{E_0}$ and $\hat{E}$ may differ by tens of degrees. For an asteroid with heterogeneous surface reflectance features, $\Delta m_B$ is strongly correlated with $\hat{\bar{E}}$. For asteroids with heterogeneous coloring, $\Delta m_B$ is more strongly correlated with $\hat{\bar{E}}$ in some bandpasses than others. In other words, the shape of the light curve (brightness measured by an observer as a function of illumination and viewing direction) of an asteroid with surface color variability is band dependent. This principle forms the basis of the statistical test for surface color heterogeneity.

We note that $\hat{\bar{E}}$ is a time-dependent three-component vector in the rotating $x, y, z$ coordinate frame of the asteroid. Since heterogeneities on the asteroid's surface are not necessarily aligned with any of these coordinate axes, rather than examining the correlation of $\Delta m_B$ with $\hat{\bar{E_{x}}}, \hat{\bar{E_{y}}},\hat{\bar{E_{z}}}$, we instead first identified the direction of maximum variation using a multiple linear regression analysis, as implemented in \textit{scikit-learn} \citep{scikit-learn}. Given a multivariate dataset of independent variables $X$ and dependent variable $Y$, we identified the linear transformation of variables in $X$ that explains the maximum amount of variation in $Y$. Applying regression analysis to the independent variables $\hat{\bar{E_{x}}}, \hat{\bar{E_{y}}},\hat{\bar{E_{z}}}$ and the dependent variable $\Delta m(\hat{\bar{E}})$, we obtained a relationship of the form $\Delta m = k_x \hat{\bar{E_{x}}} + k_y \hat{\bar{E_{y}}} + k_z \hat{\bar{E_{z}}}$, with $k_{x, y, z}$ each corresponding to a constant weighting factor. The unit vector $\hat{k} = \widehat{(k_x, k_y, k_z)}$ points along the direction of maximum variation in the rotating coordinate frame of the asteroid, thus identifying the location on the surface of the asteroid where the homogeneous model performs the worst, that is, where heterogeneity is likely to be located. This vector can be used to target that location for follow-up observations. We then applied the linear transformation $\mathcal{L}(\hat{\bar{E}}) = \hat{\bar{E}} \cdot \hat{k}$ to obtain, for each observation, a predictor variable, $\mathcal{L}(\hat{\bar{E}})$. Geometrically, $\mathcal{L}(\hat{\bar{E}})$ corresponds to the length of the projection of $\hat{\bar{E}}$ along the direction of maximum variation $\hat{k}$ for each observation.

Having thus identified the surface location with the greatest deviation between the photometric model and the full set of observations and constructing $\mathcal{L}$, we next investigated if the magnitude of this deviation depends on wavelength. For this purpose, we evaluated
for each observed wavelength band, $B$, the Spearman rank correlation coefficient, $\rho_B$, between $\mathcal{L}(\hat{\bar{E}})$ and $\Delta m_B(\hat{\bar{E}})$ for only those observations obtained within bandpass $B$. For two different wavelength bands, $B_1$ and $B_2$, the significance in units of standard deviations (z-score) of the difference between $\rho_{B_1}$ and $\rho_{B_2}$ can be evaluated using

\begin{equation}
    z_{B_1, B_2} = \frac{z_{B_1} - z_{B_2}}{\sqrt{\frac{1}{n_{B_1} - 3} + \frac{1}{n_{B_2} - 3}}},
\end{equation}

where $n_B$ is the number of observations in each band and $z_B$ is the Fisher z-transformation of the correlation coefficient $\rho_B$:

\begin{equation}
    z_B = \frac{1}{2}\ln \left( \frac{1 + \rho_B}{1 - \rho_B} \right)
\end{equation}

as in, for example, \cite{ramseyer-1979}. A significantly high $z_{B_1, B_2}$ score indicates that the difference in the correlation coefficients of the two bandpasses is highly unlikely to arise from the assumed uniformly colored model. In this work, we adopted a minimum threshold of $z_{B_1, B_2} > 5$ to identify datasets with potential evidence of surface heterogeneity, corresponding to a 5$\sigma$ detection threshold. Note that the significance test here assumes that the two correlation coefficients being measured arise from two independent samples. While in general we may expect brightnesses in one photometric band to be correlated with brightnesses in a nearby photometric band at any given time, because the measurements in each band are non-contemporaneous (i.e., they are sampled at different times), the two samples being tested arise from independent samples. In the case of contemporaneous color measurements, surface heterogeneities can be detected by directly analyzing the color as a function of illumination and viewing geometry, as in \cite{Lacerda-2009}, for example.

\begin{figure}[ht!]
    \centering
    \includegraphics[width=1\linewidth]{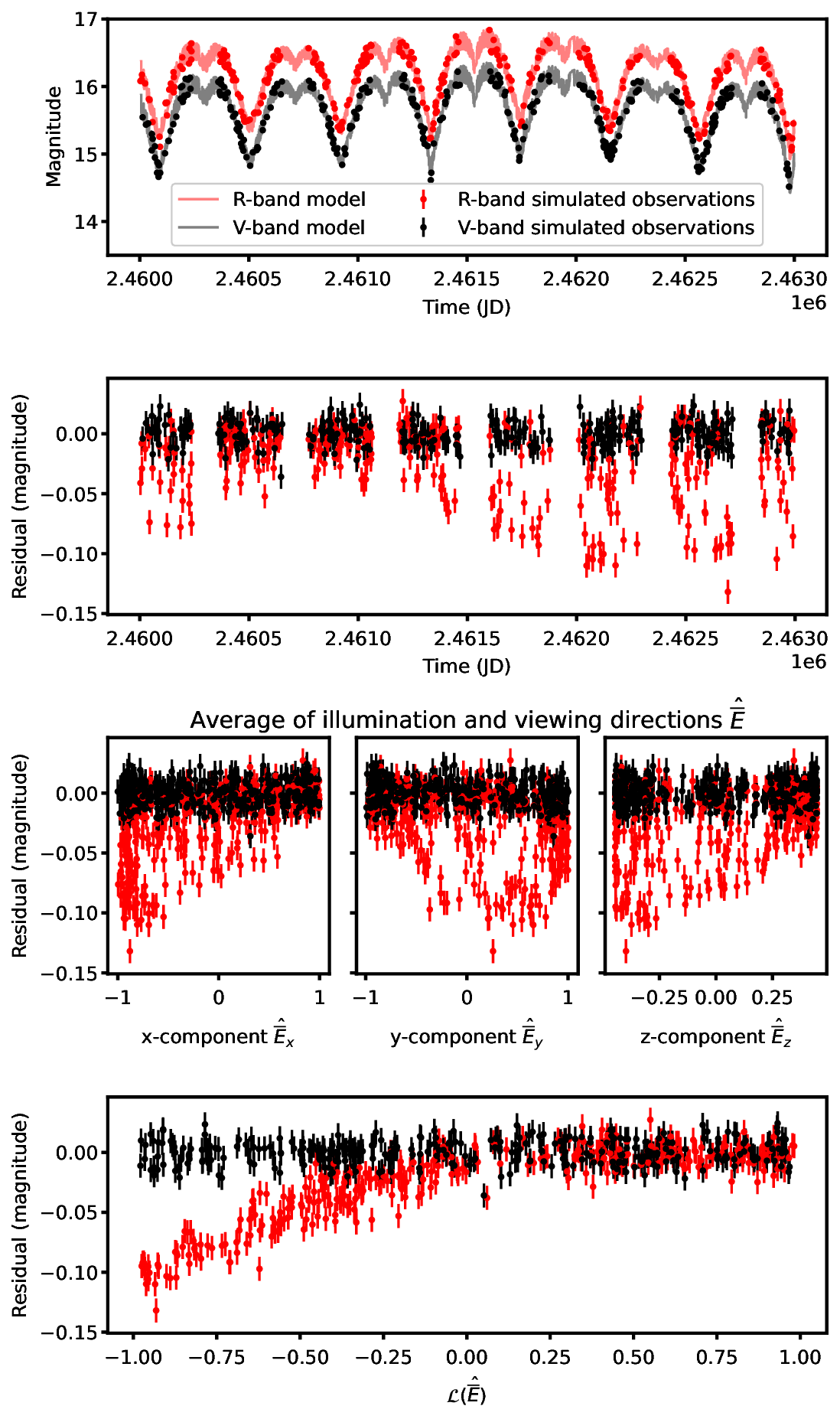}
    \caption{Graphical summary of the steps of the statistical test for surface color variation for a simulated asteroid (see Section 3 and Appendix A for a description of the photometric models) in the V and R bands. The simulated asteroid contains a large high albedo spot visible in the R band. First, a uniformly colored photometric model was fit to the simulated photometry. Second, the residuals ($\Delta m_B$) in predicted magnitude between the photometric model and the simulated photometry were computed. At this point, there were obvious discrepancies between the performance of the model in predicting the V-band and R-band, but it was unclear whether these discrepancies are associated with the viewing geometry of the asteroid. Third, the residuals in magnitude were plotted as a function of the $x,y,z-$components of the subsolar direction in the rotating reference frame of the asteroid. Fourth, multiple linear regression was used to identify the linear transformation $\mathcal{L}$ of $\hat{\bar{E_x}}, \hat{\bar{E_y}}, \hat{\bar{E_z}}$ to a single predictor variable $\mathcal{L}(\hat{\bar{E}})$ that maximizes the correlation between $\mathcal{L}(\hat{\bar{E}})$ and $\Delta m$. The final step of the test (not pictured) was to compute the significance of the difference in the correlation coefficient between $\mathcal{L}(\hat{\bar{E}})$ and $\Delta m_B$ for the two bands using the Fisher z-transformation.}
    \label{fig:test}
\end{figure}

To summarize, the steps for testing a given a dataset of multiband photometric measurements of an asteroid for heterogeneity in surface color are as follows:

\begin{itemize}
    \item Develop a photometric model for the asteroid, including a shape model, rotational pole orientation, rotational period, and bandpass-dependent phase functions under the assumption of spatial uniformity in scattering properties. 
    \item Compute the residuals $\Delta m_B(t)$ as the difference between the observed and predicted brightness of the asteroid for each observational epoch.
    \item Use ephemeris data and the rotational properties of the model to compute the average of the illumination and viewing directions $\hat{\bar{E}} = \frac{1}{2}\left(\hat{E_0} + \hat{E}\right)$ relative to the rotating coordinate frame of the asteroid for each observational epoch.
    \item Apply multiple linear regression to paired datasets $\hat{\bar{E}}$ and $\Delta m_B$, using photometry from all bandpasses to determine the linear transformation that maximizes the correlation between viewing direction and residual magnitude. Identify the direction of maximum variation $\hat{k}$ and compute $\mathcal{L}(\hat{\bar{E}})$ for each observation.
    \item For each bandpass, compute the correlation coefficient between $\mathcal{L}(\hat{\bar{E}})$ and $\Delta m_B(\hat{\bar{E}})$. Apply the Fischer-z transformation to assess whether differences in correlation coefficient between bandpasses are statistically significant.
    
\end{itemize}
The first four steps are also summarized graphically in Figure \ref{fig:test} for a simulated asteroid.

\section{Generation of synthetic asteroid photometry}

To quantify how accurately various input parameters of the asteroid photometric model must be known in order to reliably detect surface heterogeneity, we produced a set of simulated observations of 1000 synthetic asteroids with a variety of known, predetermined shapes, rotation states, and wavelength-dependent phase functions. For each synthetic asteroid, we created two sets of synthetic photometry: one for the case of a ``spotted'' asteroid with surface heterogeneity apparent in one of two observed wavelength bands and one for the case of a uniformly colored asteroid. We then computed the result of the test for surface heterogeneity assuming a ``perturbed'' photometric model, which differs from the truth model used as input to generate the synthetic photometry in a single parameter, the value of which is drawn from a (typically Gaussian) probability distribution with a mean equal to the true value of the parameter and standard deviation $\sigma$. By observing how the spot detection (correct detection of a spot in the photometry of a ``spotted'' asteroid) and false positive rates (rate of incorrectly detecting a spot in the photometry of uniformly colored models) across all 2000 synthetic datasets change as the value of $\sigma$ increased, we quantified the required accuracy for each parameter in the photometric model using a Monte Carlo approach. For each synthetic asteroid, a total of 73 perturbed models, differing from the truth model in a single parameter, were generated.

Our general approach to the photometric modeling of synthetic asteroids closely followed the general technique used to compute the direct problem of modeling the light curves of convex objects employed to solve the convex inversion problem \citep{Kaasalainen-2001, Kaasalainen-Torppa2001, Durech-2010}. A similar approach to producing synthetic light curves from a discretized triangular mesh is described in \cite{Lu-2014}.  We included minor modifications to these methods in our model to account for multiband observations and spatial variations in photometric properties, as detailed in the Appendix. To summarize, the following inputs to the model were required to obtain synthetic photometry at a set of times $t$ in wavelength bands $B$:
\begin{itemize}
    \item For each time of interest $t$, the light time corrected asteroid-observer and asteroid-Sun vectors in ecliptic coordinates (to compute the asteroid's reduced magnitude, $E,$ and $E_0$).
    \item The rotational state of the asteroid (including the ecliptic coordinates of the rotational pole, rotational period, initial rotation angle, and reference epoch) to convert the asteroid-observer ($E$) and asteroid-Sun vectors ($E_0$) from ecliptic coordinates to the rotational frame of the asteroid.
    \item A convex triangular mesh representing the shape model of the asteroid.
    \item A disk function (scattering law) $S(\mu_0(t), \mu(t))$ describing the dependence of flux on the incidence and exitance angles $\mu_0$ and $\mu$.
    \item For each wavelength band of interest $B$,
        \begin{itemize}

            \item An absolute magnitude $H_{0,B}$ representing the observed magnitude under idealized conditions with both the subsolar and sub-observer points located at ($0^\circ, 0^\circ$) longitude and latitude,
            \item The wavelength-dependent phase function $f_B(\alpha)$, 
            \item A reflectance map assigning a reflectance factor $R_T$ to each facet of the shape model.

        \end{itemize}     
\end{itemize}

In generating ephemerides for the synthetic asteroids, we selected 1000 asteroids each at random from the populations of MBAs, NEAs, and TNOs with $H_{mag} < 15$ as retrieved using the JPL Horizons database query. We aimed to evenly sample the actual distributions of orbital elements of these populations. We used the \textit{kete} \citep{kete-2025} package to propagate the asteroids' orbits and compute the heliocentric and geocentric distances $r$ and $\delta$, as well as $E$ and $E_0$ for an Earth-based observer for a total of 300 randomly sampled dates per wavelength band over the course of 3000 days. To simulate the solar elongation restraints of a ground-based telescope, we selected dates only among those intervals where the elongation of the asteroid as viewed from Earth exceeded 45$^\circ$. Rotational pole orientations were chosen from the uniform distribution on a sphere. Rotational periods from a uniform distribution ranging from 0.1 - 2 days represented the majority of asteroids rotating slower than the spin barrier \citep{Warner-2009} while excluding ultra-slow rotators, which are likelier to be tumblers in more complex rotational states \citep{Zhou-2025}. The initial rotation angle relative to the coordinate system of the asteroid was randomized between $0 - 360^{\circ}$ at the reference epoch. We modeled the shape of each asteroid as a triangulated mesh approximating an ellipsoid with asymmetric semiaxes, or Cellinoid \citep{Cellino-1989, Lu-2014} with 2000 total triangular faces. The ellipsoid's unequal semiaxes were  $a_1, a_2, b_1, b_2, c_1, c_2$, where $c_1$ and $c_2$ were the axes aligned with the asteroid's rotational pole. We assigned the longest axis to be the non-rotational axis $a_1 = 1$ and measured the lengths of the other five semiaxes relative to the longest half-axis. The direction of $a_1$ also defined the zero longitude point of the rotating coordinate axis in the frame of the asteroid. We selected the lengths of the remaining semiaxes from a uniform random distribution from 0.33 - 1 to represent a range of asteroid shapes from near-spherical to extremely oblate, with flattening observed along the axis of rotation, as would be expected of a rotating body in hydrostatic equilibrium. For the disk function, we used the Lommel–Seeliger law, an appropriate model for dark asteroid surfaces with limited multiple scattering \citep{Muinonen-2015}. 

\begin{figure}
    \centering
    \includegraphics[width=0.75\linewidth]{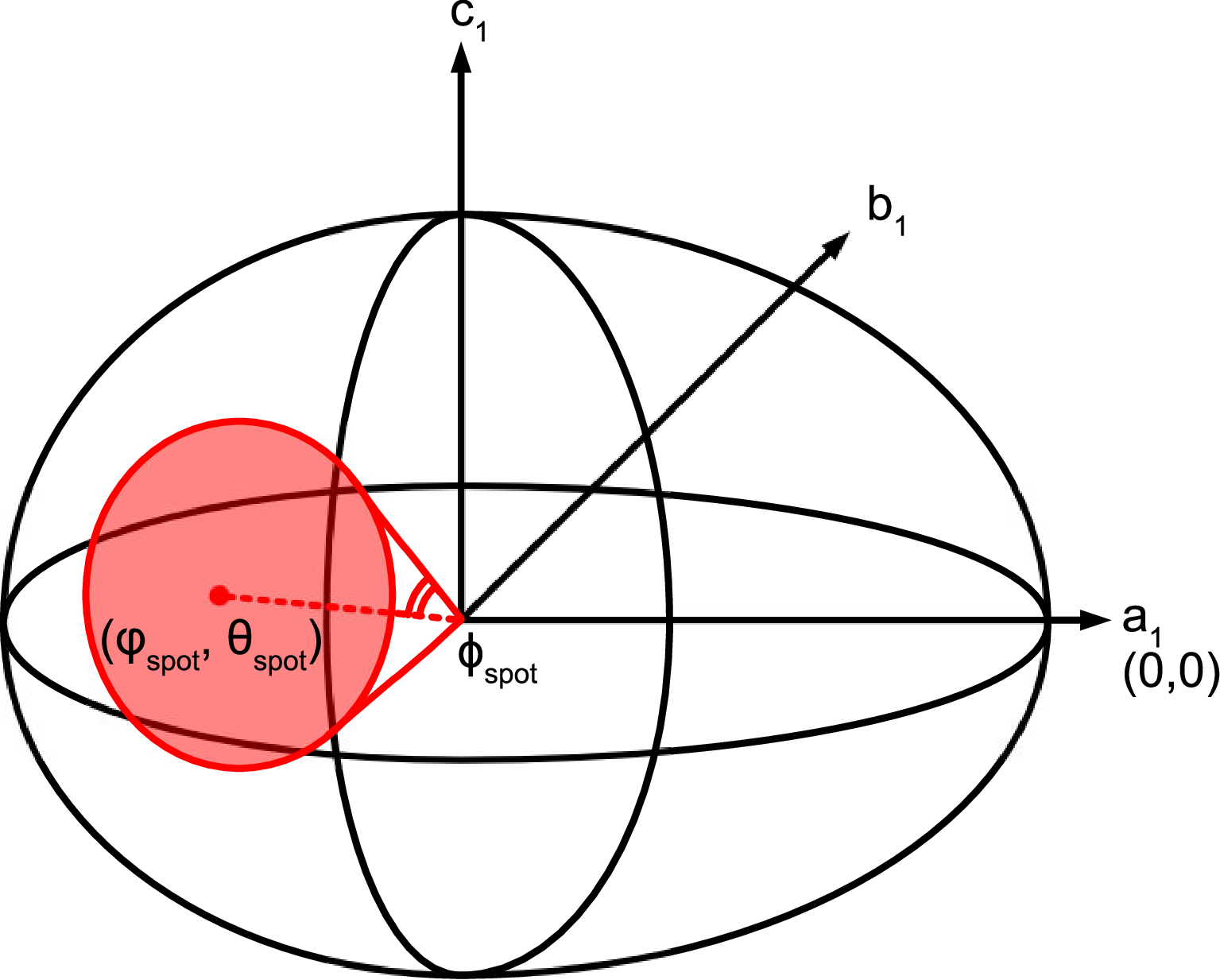}
    \caption{Parameterization of the synthetic asteroids. Each asteroid is modeled as an ellipsoid with unequal semi-axes. Semi-axis $c_1$ is aligned to the rotational pole, and the coordinate system is defined with respect to the longest semi-axis, $a_1$. Color heterogeneities (red) are parameterized by their central latitude and longitude $(\varphi_{spot}, \theta_{spot})$ and cone vertex angle $\phi_{spot}$, a measure of the angular size of the spot.}
    \label{fig:placeholder}
\end{figure}

As the statistical test deals with differences in magnitude, the value of the $H_{0, B}$ magnitude for each band was set to 0. Any deviations of the assumed value of $H_{0, B}$ from its true value manifests as a constant offset to all residuals independent of viewing direction and therefore do not affect the value of the test statistic. For simplicity, in this paper, we adopted the $HG^*_{12}$ system, which is most suitable for sparse datasets and minimizes the number of free parameters while taking into account the observed correlation between $G_1$ and $G_2$ in more complex models \citep{Penttila-2016}. We selected the value of $G^*_{12}$ for each band from a uniform random distribution from 0 - 1. 

For the reflectance maps in each band, we assumed a constant reflectance map $R_T = 1$ in both wavelength bands to produce the uniform model. For the spotted model, we adopted a simple model of an asteroid with constant reflectance in the first band and a circular spot of enhanced reflectance in the second band. The spot was parameterized by the position of the center of the circular spot (latitude $\varphi_{spot}$ and longitude $\theta_{spot}$) in the rotating coordinate frame of the asteroid, the angular size of the spot (parameterized as the vertex angle $\phi_{spot}$ of a cone pointing outward toward $\varphi_{spot}, \theta_{spot}$ from the asteroid's center), and a reflectance enhancement factor $R_{f}$. All triangular facets of the shape model with centers located within the cone were assigned reflectances of $R_T = R_{f}$, while those outside the cone were assigned reflectances $R_T = 1$. In other words, triangle $T$ with center $T_x, T_y, T_z$ in the rotating coordinate frame of the asteroid was assigned a reflectance $R_f$ if the dot product of $(T_x, T_y, T_z)$ and the direction pointing from the center of the asteroid to the center of the spot exceeded $\cos \phi_{spot}$. A cone with $\phi_{spot} = 90^{\circ}$ corresponded to a hemisphere-sized spot. The central latitude and longitude of the spot was selected by choosing a direction from the uniform random distribution over the sphere, $\phi_{spot}$ from a uniform random distribution from $20 - 90^{\circ}$, and $R_f$ from a uniform random distribution from 1.2 - 2. Note that by this definition, spots are always brighter than their surroundings. We did not find a significant difference between the detection and false positive rates for dark and bright spots with the same color contrast relative to their surroundings; for example, the detection rates were equivalent for spots with reflectance enhancements of $R_f$ and $1/R_f$. To reduce redundancy, we present the results of each test for only the ``bright'' spot case. In reality, the distribution of color on asteroid surfaces is likely to be more complicated than a single, uniform spot. Disk integrated photometry is inherently sensitive to variations in average color at large regional scales only, resulting in a fundamental limitation on the size of color variations detectable based on unresolved photometry alone. Therefore, for the present study we limited our scope to a single, relatively large region of enhanced average flux.

For all sensitivity tests, with the exception of the sensitivity to random measurement error test in Section 4.2, we added Gaussian noise with a standard deviation of 0.01 magnitudes to the synthetic magnitudes predicted by the perturbed photometric model to simulate the effect of inherent random measurement errors. This level of random measurement error was chosen to be comparable to the expected upper limits on photometric error for the LSST and ATLAS surveys \citep{Tonry-2018, LSST-2019}.

\section{Performance of the test}

\subsection{Application of the test to different small body subpopulations}

\added{We evaluated the effect of observing small bodies in different dynamical populations (MBA, NEA, and TNO) on the spot detection rate. We computed $z_{B_1, B_2}$ for observations of objects with different orbital parameters. Using a detection threshold of 5$\sigma$, we computed the true detection and false positive rate for the spotted and uniform synthetic asteroids, respectively.}

\begin{table}[]
    \caption{Detection rates per small body population as a function of spot size.}
    \centering
    \begin{tabular}{c|c|c|c}
    \hline \hline
          Spot size $(^{\circ})$ & NEAs & MBAs & TNOs \\
    \hline
         $80 < \phi_{spot} \leq 90$ & 1.00 & 1.00 & 0.99 \\
         $70 < \phi_{spot} \leq 80$ & 1.00 & 1.00 & 0.98 \\
         $60 < \phi_{spot} \leq 70$ & 1.00 & 1.00 & 0.96 \\
         $50 < \phi_{spot} \leq 60$ & 1.00 & 1.00 & 0.93 \\
         $40 < \phi_{spot} \leq 50$ & 0.99 & 1.00 & 0.92 \\
         $30 < \phi_{spot} \leq 40$ & 0.99 & 0.99 & 0.85 \\
         $20 < \phi_{spot} \leq 30$ & 0.91 & 0.96 & 0.74 \\
    \end{tabular}
    \label{tab:orbit}
\end{table}

The results are shown in Table \ref{tab:orbit}. The detection rate for all spot sizes is highest for MBAs and lowest for TNOs, with NEAs achieving intermediate values. The detection rates for TNOs especially are strongly dependent on the angular size of the spot $\phi_{spot}$. The false positive rate is low ($\leq 0.1\%$) for all subpopulations. We note that the detection rate as a function of subpopulation is also dependent on survey design: as detections of color changes associated with seasonal variations in aspect angle are possible only for objects which can be observed over many seasons, the total duration of the survey limits detections of latitudinal variation in color to those objects with orbital timescales less than or equal to the total duration of the survey. This is the primary factor strongly limiting the detection rate for TNOs: the test is simply not able to detect color variations on regions of the object not visible and illuminated during the duration of the survey, and this explains why smaller spots are more likely to be missed. As a result, detections of latitudinal variation in color are limited to asteroids in the Main Belt or closer. The situation is slightly more complicated for NEAs. Due to elongation constraints, observing these asteroids at all is only possible during certain portions of their orbital period, inherently limiting the possible aspect angles at which they are able to be observed, resulting in their intermediate detection rates. Additional information (such as brightness constraints and observation cadence) would be required to assess the dependence of detection rates on orbital elements in greater detail for a particular survey strategy. Given the expected timescales for present and near-future photometric surveys are on the order of the orbital timescales of MBAs, in subsequent tests, we limited our analysis to the MBA subpopulation only in order to derive upper bounds on the detection rate as a function of each perturbed variable. However, the test for heterogeneity is still applicable to TNO and NEA populations so long as the observations in each band samples the same distribution of latitudes and longitudes, to avoid false positive detections (see discussion in next section). In this case, the test is only sensitive to heterogeneities within the common range of latitudes and longitudes sampled by both bands.

\subsection{Minimum number of observations}

We evaluated the effect of varying the total number of observations of a given asteroid on the detection rate of surface spots. First, we assumed each band was observed an equal number of times. We computed $z_{B_1, B_2}$ for a subset of the synthetic observations. Using a detection threshold of 5$\sigma$, we computed the true detection and false positive rate for the spotted and uniform synthetic asteroids respectively as a function of the number of times the asteroid was observed.

\begin{figure}[ht!]
    \centering
    \includegraphics[width=1\linewidth]{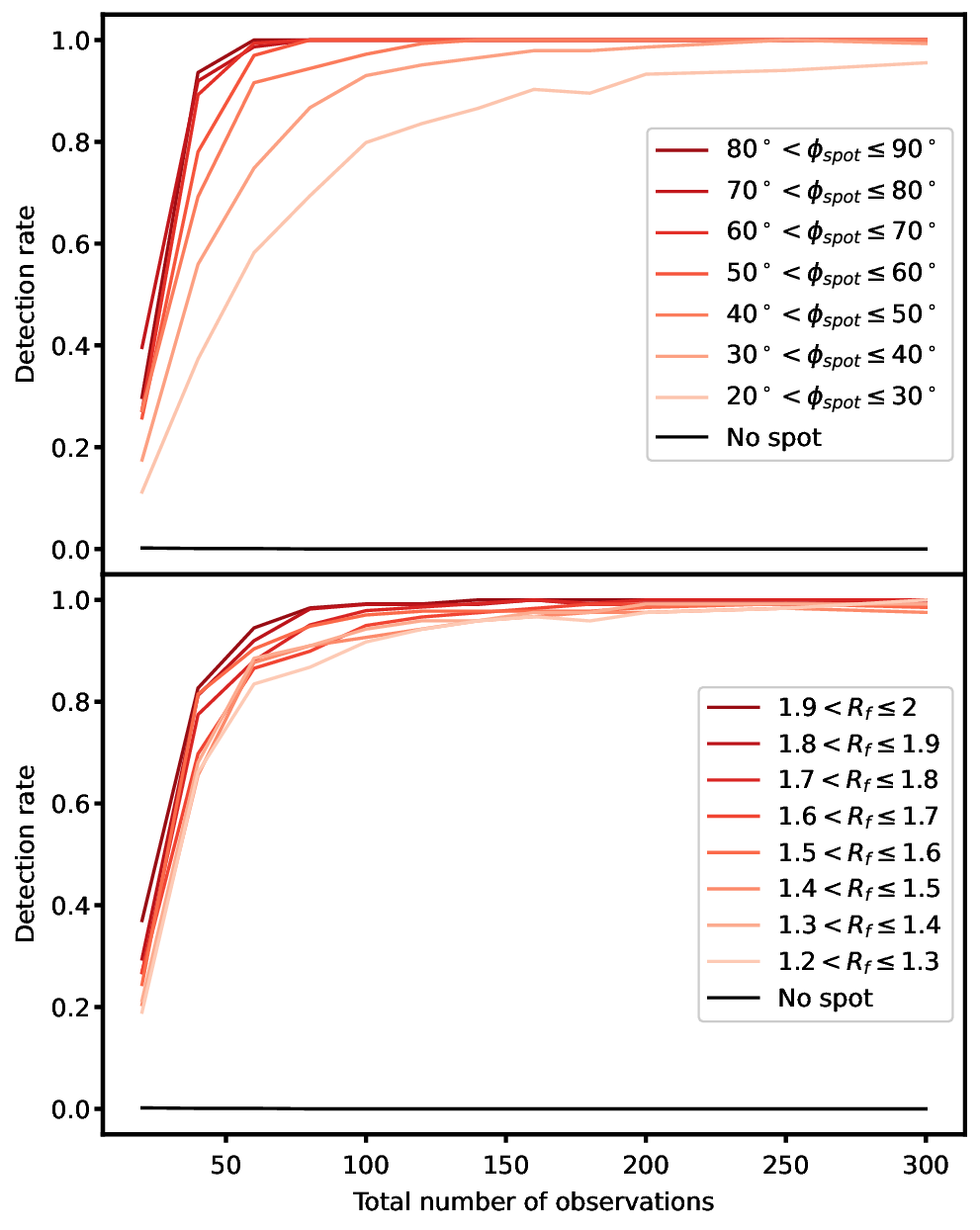}
    \caption{Detection and false positive rates as a function of the number of times an asteroid is observed. Detection rates depend on spot properties, including the angular size of the spot $\phi_{spot}$ and the reflectance enhancement factor within the spot $R_f$.}
    \label{fig:nobs_test}
\end{figure}

The results are shown in Figure \ref{fig:nobs_test}. For all spot sizes and reflectance enhancement factors, the detection rate increases as more observations are taken. Spots with smaller angular sizes are more difficult to detect with a small number of observations, as the likelihood of failing to observe the asteroid when a small spot is illuminated and visible is relatively high when a limited number of observations are available. The false positive rate is $<1\%$ and decreases as more observations are taken. Based on the results, we estimate a conservative minimum of 100 - 150 observations total are needed to detect 90\% of spots (depending mainly on spot size). 

\begin{figure}
    \centering
    \includegraphics[width=0.75\linewidth]{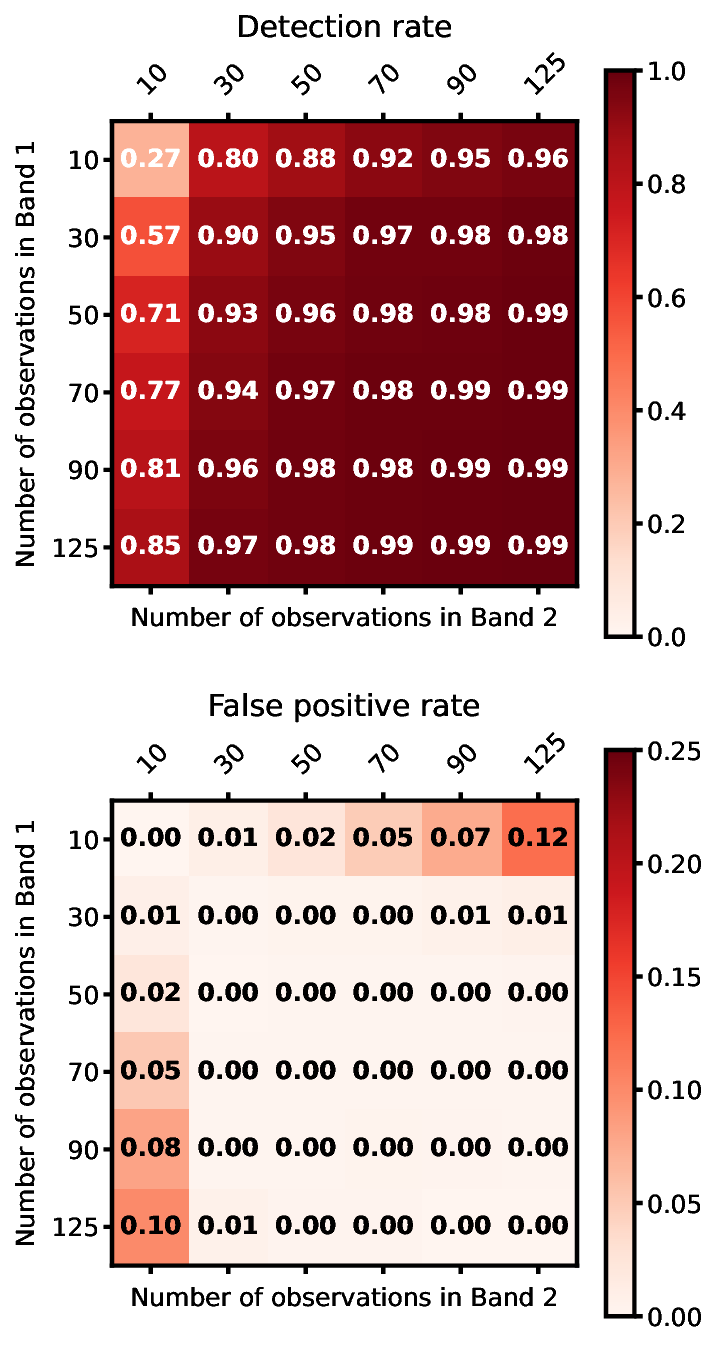}
    \caption{Detection and false positive rates as a function of the number of observations in each of the two bands. The synthetic asteroid has a uniform albedo in Band 1 but displays an albedo spot in Band 2.}
    \label{fig:nobs2_test}
\end{figure}

While we assumed the asteroid would be observed an equal number of times per bandpass, this is likely to not be the case for real data. Next, we evaluated the effect of varying the number of observations in each band independently. The results are shown in Figure \ref{fig:nobs2_test}. In general, the detection rate increases as the number of observations in either band increases. However, there is a slight asymmetry: increasing the number of observations in Band 2 increases the detection rate more than increasing the number of observations in Band 1 does. This is because the spot is only present in Band 2: the reflectance in Band 2 is not uniform as it is in Band 1, so the uniform albedo model performs more poorly for Band 2. The discrepancy between the model and observations is more readily apparent as more Band 2 observations are included. With real observations, we do not know a priori which band an albedo spot is present in, if at all, so the band that the uniform model performs best for will be dominated by the band with the most observations and/or the smallest error bars. Therefore the detection rates for the case where the number of observations in Band 1 exceeds the number of observations in Band 2 provide a more realistic estimate of the detection rate for situations in which an asteroid is not observed an equal number of times in each bandpass.

The performance of the test with low numbers of observations, especially the elevated false positive rate when only one band is well sampled ($>30$ observations), illustrates an additional criterion for the test's application. In addition to requiring a minimum number of observations, the distribution of subsolar and sub-observer latitudes and longitudes sampled by each band should be similar. Potential false positives can arise if a region of the asteroid is particularly poorly modeled by the assumed photometric model and this region is only (or predominantly) sampled in only one band. In this case, without sufficient data for the problematic region in the second band, it is impossible to distinguish between poor model performance due errors in assumed parameters (e.g., an inaccurate shape model) versus true surface color heterogeneity. When a single band only has a few observations, it is more likely that those observations do not cover the same range of subsolar and sub-observer latitudes and longitudes sampled by the more well observed band. While sampling the full range of possible subsolar and sub-observer latitudes and longitudes improves overall detection rates, for some objects, particularly those beyond the Main Belt, obtaining such observations would require observations spanning infeasibly long timescales. However, the test can still be applied by restricting the test to those photometric data that sample a common distribution of latitudes and longitudes.

\subsection{Sensitivity to random measurement error}

We evaluated the effect of adding random Gaussian noise to the 
simulated magnitudes to approximate the effect of measurement errors arising from real-world observing conditions. We computed $z_{B_1, B_2}$ for observations with increasing levels of noise in measured magnitudes, assuming all other model parameters had been accurately inferred. Using a detection threshold of 5$\sigma$, we computed the true detection and false positive rate for the spotted and uniform synthetic asteroids respectively as a function of the standard deviation of the added Gaussian noise. 

\begin{figure}[ht!]
    \centering
    \includegraphics[width=1\linewidth]{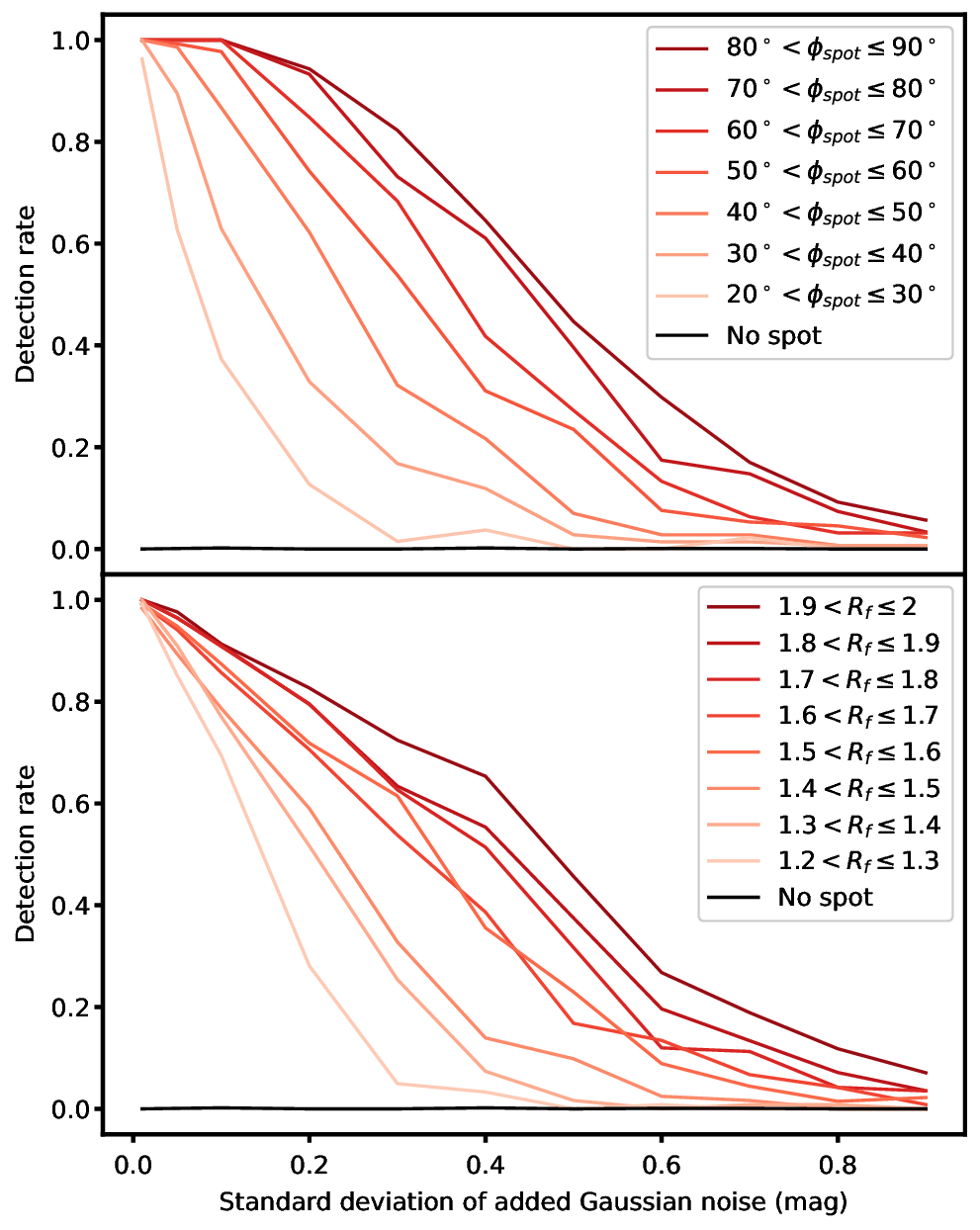}
    \caption{Detection and false positive rates as a function of increased added Gaussian noise. Detection rates depend on spot properties, including the angular size of the spot ($\phi_{spot}$) and the reflectance enhancement factor within the spot ($R_f$).}
    \label{fig:noise_test}
\end{figure}

The results are shown in Figure \ref{fig:noise_test}. For all spot sizes and reflectance enhancement factors, the true detection rate decreases to near 0 as the standard deviation of the added noise increases. The false positive rate remains near 0 for all noise levels. Adding random Gaussian noise to the measured magnitude increases scatter among the model residuals in the $\Delta m_B$ direction, without any dependence on $\hat{\bar{E}}$. The added scatter effectively reduces the measured correlation coefficient between $\hat{\bar{E}}$ and $\Delta m_B$ in bands where a correlation is present without introducing spurious dependencies on $\hat{\bar{E}}$, resulting in the observed dependence of the detection rate and the independence of false positive rate on the level of increased noise. Spots that are larger in size or with stronger flux enhancements are more readily detectable at high noise levels. The detection rate of spots with widths less than $20-30 ^{\circ}$ drops dramatically as a function of noise, reaching $<50\%$ for observations with $>0.1$ mag uncertainties. The $\phi_{spot}>20^{\circ}$ limit gives a realistic lower bound on the minimum angular size of regional color variations that could be inferred from whole-disk photometric measurements. Similarly, the detection rate of spots with reflectance enhancement factors of $R_f<1.2$ is strongly affected by relatively low levels of added noise and sets a conservative lower bound on the strength of the color differences likely to be detectable. For any given survey strategy and instrument, the availability of a sufficient number of measurements with sufficiently low photometric noise will determine the limiting magnitude of the asteroids to which the test can be appropriately applied.

\subsection{Sensitivity to error in assumed rotational period}

We evaluated the effect of assuming an incorrect rotational period in the fit photometric model. Assuming all other model parameters were accurately inferred, we adopted a period estimate drawn from a Gaussian distribution with mean equal to the true period with a variable standard deviation. Using a detection threshold of 5$\sigma$, we computed the true detection and false positive rate for the spotted and uniform synthetic asteroids respectively as a function of the standard deviation of the added uncertainty in period.

\begin{figure}[ht!]
    \centering
    \includegraphics[width=1\linewidth]{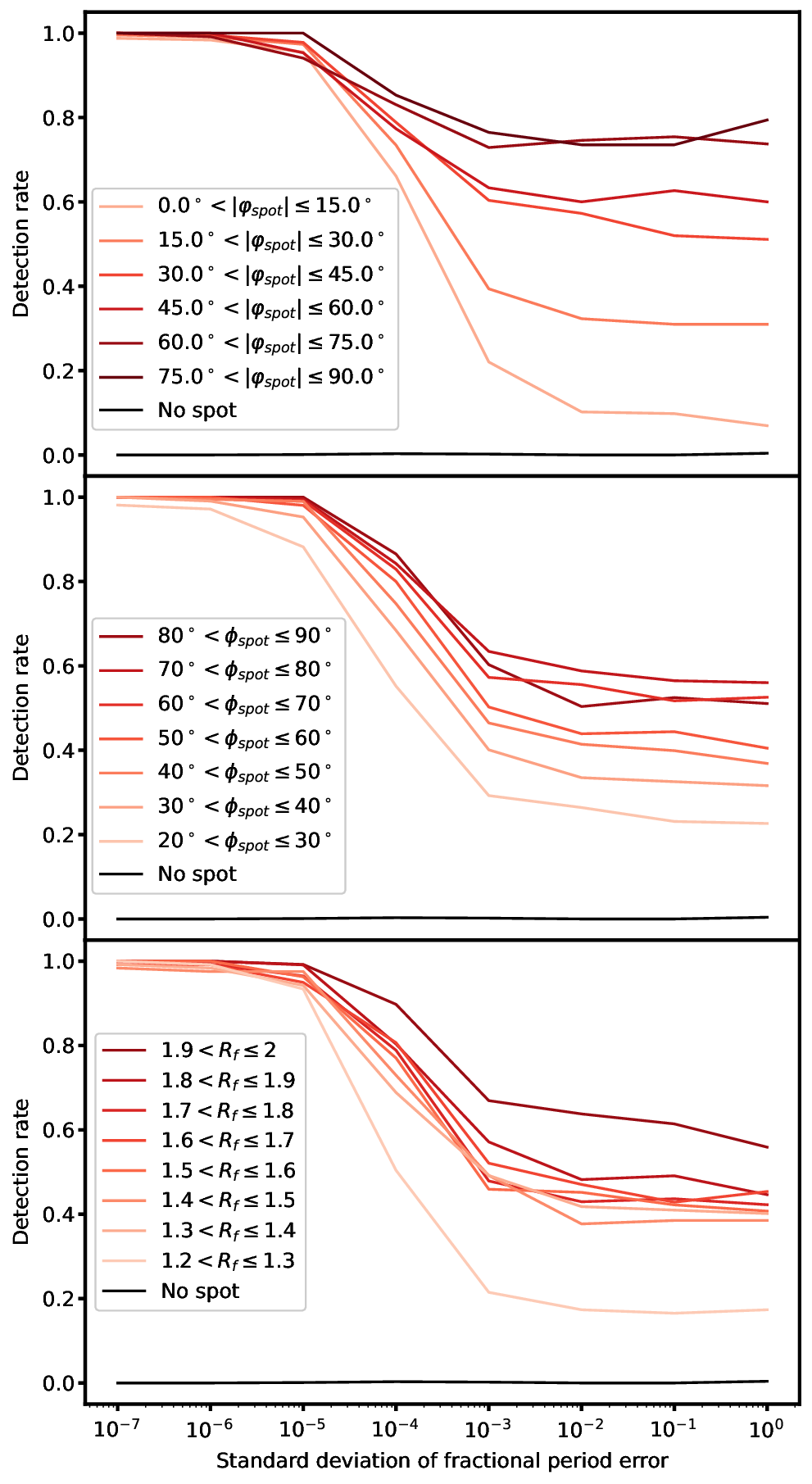}
    \caption{Detection and false positive rates as a function of increasing error in rotational period. Detection rates depend on spot properties, including the latitude of the spot ($\varphi_{spot}$), angular size of the spot ($\phi_{spot}$), and the reflectance enhancement factor within the spot ($R_f$).}
    \label{fig:period_test}
\end{figure}

The results are shown in Figure \ref{fig:period_test}. As in the previous test, the rate of false positives is not strongly dependent on error in assumed period. For all spot sizes and reflectance enhancement factors, the true detection rate decreases as the error in assumed period increases. However, the detection rate does not drop to 0, with some proportion of spots still remaining detectable even as error in assumed period grows very large. Spots with larger angular sizes and flux enhancements remain detectable at higher errors in assumed period. The detection rate at high errors in assumed period is also very strongly dependent on the latitude of the spot: for a high latitude spot, the visibility and illumination is more strongly seasonally, rather than diurnally, dependent. These polar spots remain detectable as variations in color from apparition to apparition can be readily identified without precise knowledge of the asteroid's rotational period. Adding uncertainty to the assumed rotational period effectively ``spreads out'' any longitude-dependent color variations along the longitudinal direction as the discrepancy between the assumed and true subsolar longitudes grows larger as the asteroid continues to rotate. For large enough errors, the relationship between predicted and true subsolar longitude is essentially randomized, erasing any correlations between $\Delta m_B$ and the longitudinal component of $\hat{E}_0$ and resulting in a low detection rate for near-equatorial spots. To reliably detect color variations with strong longitudinal dependence, the rotation period of the asteroid must be known to better than one part in 10,000. 

The precision to which an asteroid's rotational period is known varies widely. \citet{Durech-2022} estimated using bootstrap methods, that, of the $\sim100,000$ sparsely observed asteroids in the ATLAS dataset with more than 100 photometric observations, the period was able to be reliably determined to within one part in 10,000 for 4243 asteroids ($\sim4\%$) from sparse data alone. Errors of less than one part in 10,000 in period are achievable for well-studied asteroids with dense light curve coverage spanning multiple epochs, (i.e., \citet{Lowry-2007,Durech-2008,Marciniak-2018}), but obtaining these data is time-intensive, and they are available for a limited subset of asteroids. Combining both dense and sparse photometric datasets to determine rotational period results in an increased confidence in rotational period determination \citep{Santana-Ros-2015,Gowanlock-2024}, and single apparition dense light curves can be obtained without the need for additional observations by using archival data in the case of serendipitously observed asteroids in wide field surveys, such as TESS \citep{Pal-2020, McNeill-2023}. Imprecise knowledge of an asteroid's rotational period derived from sparse data alone represents a major limitation for detecting longitudinal surface color variability from sparse datasets, limiting the utility of the test to near-polar spots for most asteroids. However, the required 1 in 10,000 precision in rotational period can be achieved using sparse data alone for $\sim4\%$ of asteroids, in addition to the subset of asteroids with dense light curves available as supplementary datasets.

\subsection{Sensitivity to error in assumed rotational pole orientation}

We evaluated the effect of assuming an incorrect rotational pole orientation in the fit photometric model. Assuming all other model parameters were accurately inferred, we adopted a pole orientation from the von Mises-Fisher distribution with mean direction equal to the true pole solution and variable concentration parameter $\kappa$. The von Mises-Fisher distribution is analogous to the Gaussian distribution in directional space, with concentration parameter $\kappa$ analogous to the reciprocal of the variance $1/\sigma^2$. At large values of $\kappa$ the spread of angles is smaller, and its distribution tends toward a multivariate Gaussian distribution with standard deviation $\sigma \approx \sqrt{1/\kappa}$ \citep{Fisher-1953}. Using a detection threshold of 5$\sigma$, we computed the true detection and false positive rate for the spotted and uniform synthetic asteroids respectively as a function of the concentration parameter of the added uncertainty in pole orientation.

\begin{figure}[ht!]
    \centering
    \includegraphics[width=1\linewidth]{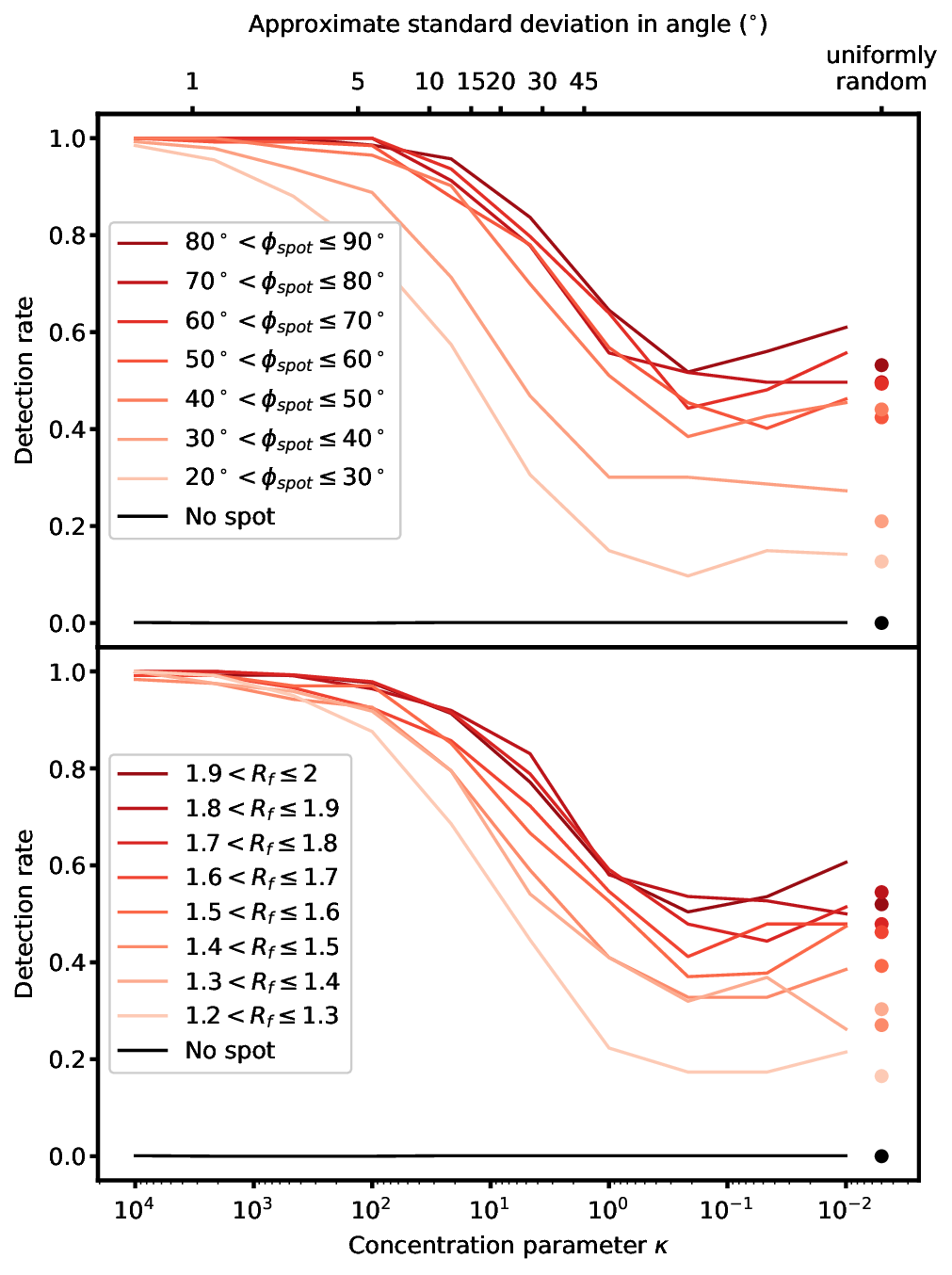}
    \caption{Detection and false positive rates as a function of increased error in assumed pole orientation. The error in pole orientation was parameterized by the concentration parameter $\kappa$ of the von Mises-Fisher distribution. A corresponding approximate standard deviation in angle (in degrees) is included for reference in the upper panel. Detection rates are also shown for the case of a randomly chosen assumed pole orientation. Detection rates depend on spot properties, including the angular size of the spot ($\phi_{spot}$) and the reflectance enhancement factor within the spot ($R_f$).}
    \label{fig:pole_test}
\end{figure}

The results are shown in Figure \ref{fig:pole_test}. As with previous tests, the false positive rate is low and not strongly dependent on error in assumed pole orientation. For increasing error in assumed pole orientation, the true detection rate decreases, with spots larger in size or flux enhancement achieving higher detection rates at large uncertainties in pole orientation. As the distribution of assumed pole orientation approaches a uniform distribution over a sphere, the true detection rate does not fall to 0. Given that the observed color for an unresolved source is derived from the flux of individual facets integrated over a hemisphere, some uncertainty in the exact pole orientation in relationship to the location of the spot is evidently tolerable. Spots remain detectable so long as the assumed pole solution can correctly identify if some portion of the spot is lit and visible to the observer at a given point in time (this requires less precision if the spot itself is large), with higher order effects on the shape of the color-light curve arising from the spot's relative location within the visible and illuminated region. For well-characterized asteroids with both dense light curves and sparse photometric datasets, errors in pole orientation can be as low as $\sim1\%$ \citep{Muinonen-2020}. For asteroids with detailed shape models recovered via light curve inversion of sparse photometric data, a typical error in pole orientation is 5-10$^{\circ}$ \citep{kaasalainen-2002,Torppa-2003,hanus-2011}. Of the $\sim100,000$ sparsely observed asteroids in the ATLAS dataset with more than 100 photometric observations, detailed shape and spin (including rotational period and pole orientation) models were successfully derived for $\sim3\%$ \citep{Durech-2020}. However, pole orientations can also be estimated without deriving detailed shape models by assuming simplified shape approximations for asteroid shapes such as triaxial ellipsoids or oblate spheroids. Under the oblate spheroid assumption, \citet{Carry-2024} computed well-fitting pole orientations for $\sim50\%$ of Solar System objects observed by the ZTF and estimate an average error in assumed pole orientation of $20^{\circ}$ (however, the rotational periods of these asteroids were not determined via this method and would require supplementary data in order to perform the test for surface color variation). Therefore, it is possible to obtain sufficiently accurate pole orientation estimations for a large proportion of the asteroid population from sparse photometry alone--the requirement for accuracy on pole orientation alone is more relaxed than the requirement for period accuracy. 

\subsection{Sensitivity to error in assumed shape}

We evaluated the effect of assuming an incorrect shape model in the fit photometric model. Assuming all other model parameters were accurately inferred, for each half-axis ratio (e.g., $a_2:a_1, b_1:a_1, b_2:a_1,$ etc.) we adopted an assumed half-axis ratio drawn from a Gaussian distribution with a mean equal to the true half-axis ratio and a variable standard deviation. Because each half-axis ratio must be positive, any negative values generated from the Gaussian distribution were discarded and resampled until a positive value was obtained. Using a detection threshold of 5$\sigma$, we computed the true detection and false positive rate for the spotted and uniform synthetic asteroids respectively as a function of the standard deviation in assumed axis ratios. We also computed the detection and false positive rates for the sample if the assumed shape model was perfectly spherical, as would be assumed in the absence of shape information. 

\begin{figure}[ht!]
    \centering
    \includegraphics[width=1\linewidth]{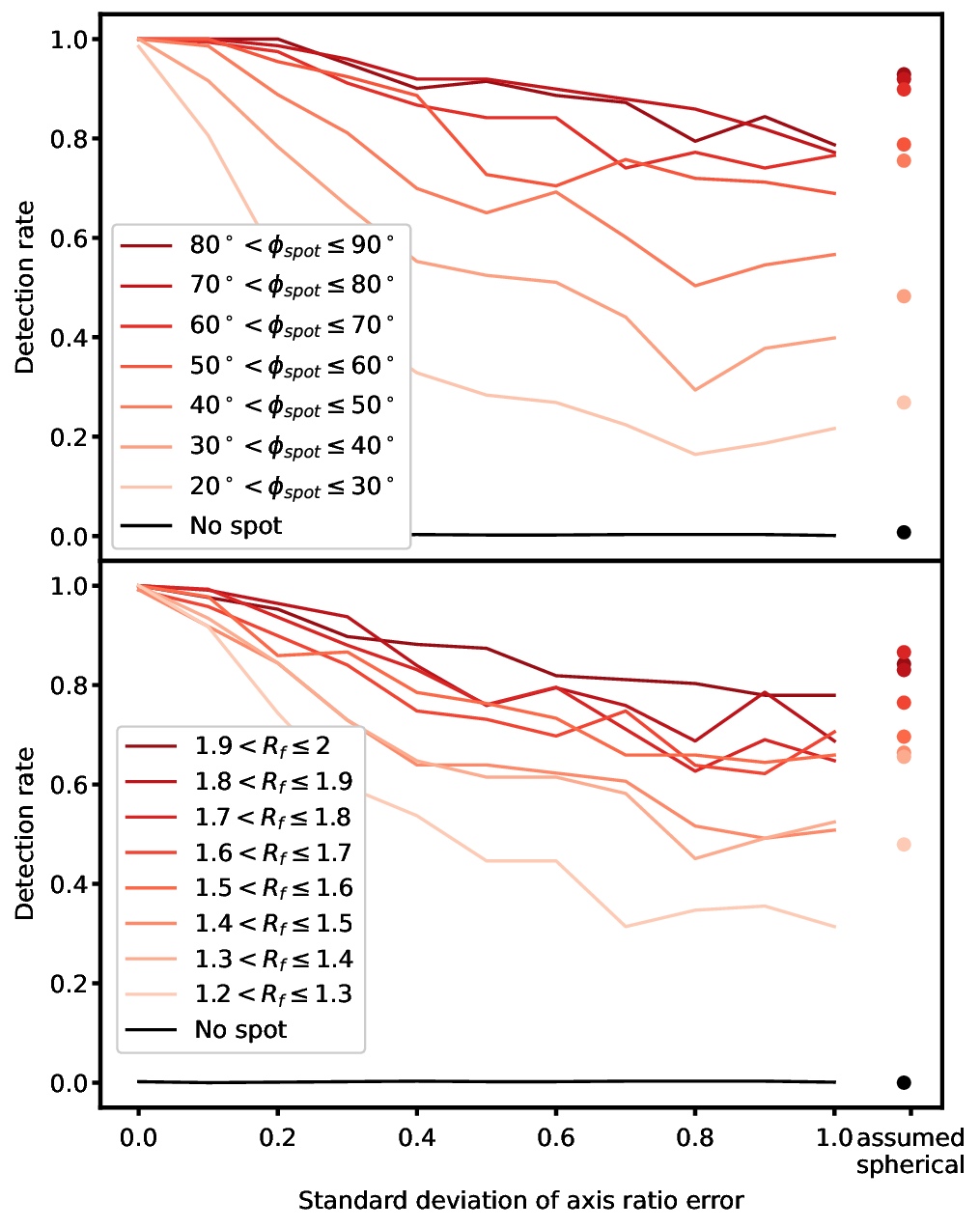}
    \caption{Detection and false positive rates as a function of error in assumed axis ratios. Detection rates are also shown for the case of an assumed spherical shape. Detection rates depend on spot properties, including the angular size of the spot ($\phi_{spot}$) and the reflectance enhancement factor within the spot ($R_f$).}
    \label{fig:shape_test}
\end{figure}

The results are shown in Figure \ref{fig:shape_test}. Similar to the previous test, the false positive rate is low and not strongly dependent on the error in assumed axis ratios. Because the same shape model is assumed for all bands, introducing an incorrect shape affects all wavelength bands to the same extent. As the error in assumed axis ratio increases, the true detection rate decreases, with spots with smaller angular sizes and flux enhancements more strongly affected by error in assumed axis ratio. Even if the asteroid is incorrectly assumed to be spherical, a large proportion of spots remain detectable. Assuming a spherical shape at some point gives better results than an erroneous shape model.

As discussed in the previous section, detailed shape models can be recovered via light curve inversion for $\sim3\%$ of asteroids with $>100$ sparse photometric observations \cite{Durech-2020}. Shape models derived from light curve inversion are estimated to include errors in axis ratio, particularly the b/c axes ratios, of up to 10\% \citep{Torppa-2003}. As with pole orientation, approximate shapes can be derived for a larger number of asteroids by assuming a simplified ellipsoidal or spheroidal shape model, at the cost of increasing the uncertainty in assumed axis ratio. For the asteroids well-approximated as oblate spheroids using the methods of \cite{Carry-2024}, the estimated median error in oblateness is $\sim45\%$. \footnote{Estimate retrieved using the FINK broker \citep{fink-broker}, v. 202511} As with rotational pole orientation, it is plausible to obtain sufficiently precise shape estimates from sparse photometry alone for a significant fraction of asteroids using current light curve inversion strategies.

\subsection{Sensitivity to error in assumed phase function}

We evaluated the effect of assuming an incorrect phase function in the fit photometric model. Assuming all other model parameters were accurately inferred, for each band-specific phase function $F_B(\alpha)$  we adopted an assumed phase parameter $G_{12, B}^*$ drawn from a Gaussian distribution with mean equal to the true phase parameter and variable standard deviation. Because values of $G_{12}^*$ are necessarily constrained between 0 and 1, any value generated from the Gaussian distribution falling outside the interval $0 < G_{12}^* < 1$ were discarded and resampled until a value within the interval was obtained--for large enough standard deviations, this distribution approaches the uniform distribution over the interval $0 < G_{12}^* < 1$. 

\begin{figure}[ht!]
    \centering
    \includegraphics[width=1\linewidth]{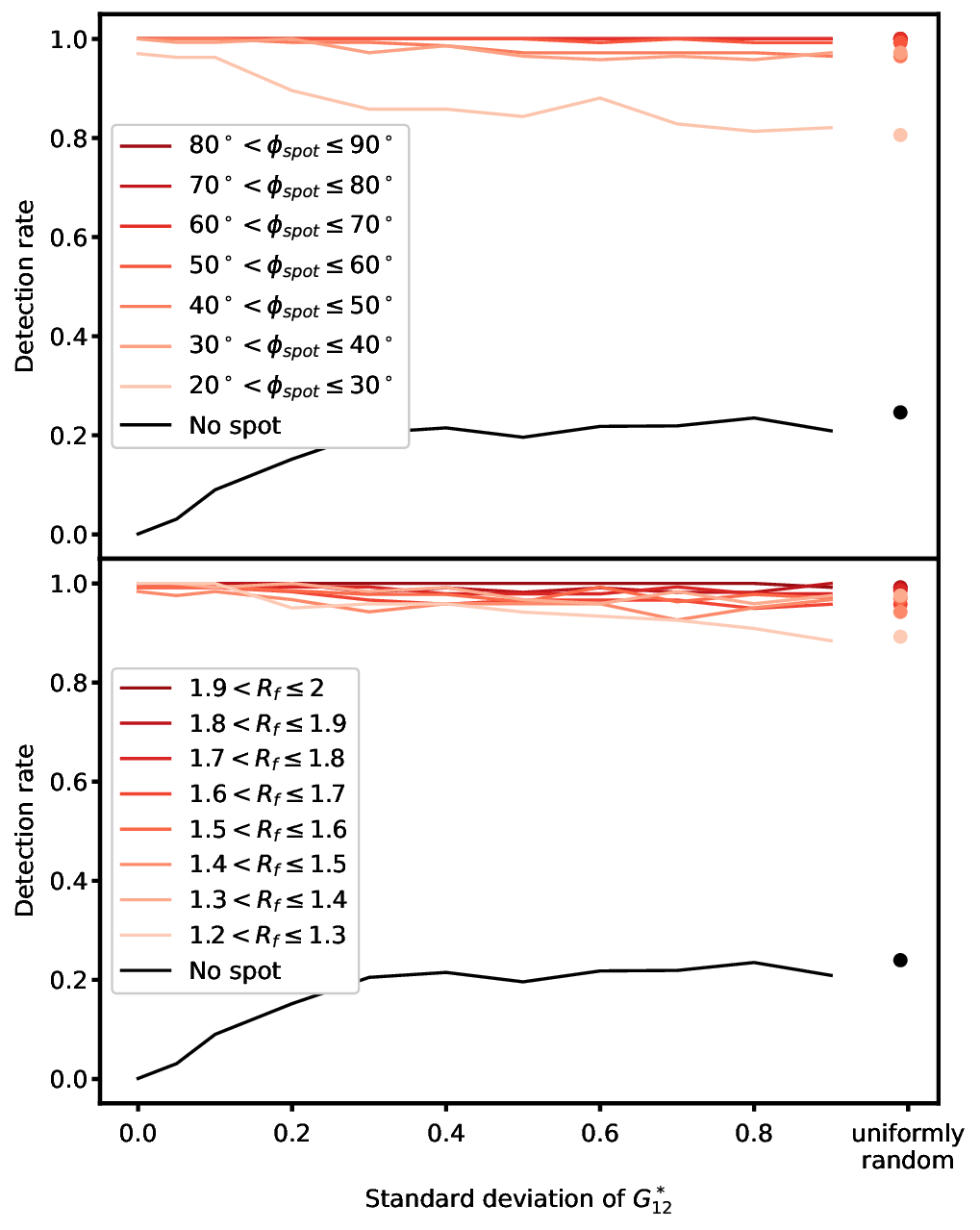}
    \caption{Detection and false positive rates as a function of increased error in the assumed phase function parameter $G_{1,2}^*$. Detection rates are also shown for the case of a uniformly randomly chosen $G_{1,2}^*$ parameter. Detection rates are reported for spots of varying properties, including the angular size of the spot ($\phi_{spot}$) and the reflectance enhancement factor within the spot ($R_f$).}
    \label{fig:g12_test}
\end{figure}

The results are shown in Figure \ref{fig:g12_test}. Unlike previous tests, the true detection rate remains at or above 80\% for spots of all sizes and flux enhancements even if the value of $G_{12}^*$ is drawn from a uniform distribution over the interval $0 < G_{12}^* < 1$. However, frequent false positive detections can arise when an incorrect value of $G_{12}^*$ is assumed, with larger errors in assumed $G_{12}^*$ resulting in a higher false positive rate. If $G_{12}^*$ is chosen from a random, uniform distribution the false positive rate is $\sim$25\%, which we estimate as the maximum false positive rate for asteroids with uncertain phase functions. \citet{Mahlke-2021} estimated the $HG_{12}^*$ parameters and their errors for over 100,000 asteroids using sparse photometry from the ATLAS survey, finding a median error in $G_{12}^*$ of 0.06. It is thus plausible to obtain phase function estimates precise enough that the expected false positive rate is below 10\% for a large fraction of the asteroid population. False positives often arise when the magnitude of the opposition surge is not well-characterized by the assumed photometric model, leading to large residuals observed at low phase angles and an apparent correlation between the strength of the model residuals and the subsolar and sub-observer locations that happened to be observed at low phase angles. More sophisticated models, such as the $HG_1G_2$ model, provide more degrees of freedom to model the strength of the opposition surge as an independent parameter, and should be considered on a per-asteroid basis to evaluate possible detections of heterogeneity. We also note that while the values of $G_{12}^*$ for each band were independently selected from a uniformly random distribution in our simulations, in reality, the value of the phase function parameter $G_{12}^*$ in one band may be correlated with its value in other bands. The results of \cite{Mahlke-2021} suggest that this is indeed the case, with $G_{12,o}^*$ and $G_{12,c}^*$ in the orange and cyan bands respectively somewhat correlated with a correlation coefficient $\sim 0.5$. However, the standard deviation of $G_{12,o}^* - G_{12,c}^*$ in the \cite{Mahlke-2021} dataset is $\sim 0.3$, indicating that assuming any particular fixed relationship between the value of $G_{12}^*$ in different bands is of limited utility in terms of decreasing the false positive rate given the required accuracy in $G_{12}^*$. Instead the phase function parameters should be independently fit on a per-band basis. Finally, phase function modeling that takes into account the effects of uneven illumination on nonspherical asteroids reduces the scatter in residuals in the phase curve, leading to more accurate estimations of phase curve parameters \citep{Carry-2024, Oszkiewicz-2025}. Obtaining more accurate phase functions via the inclusion of shape effects as part of the phase function fitting process reduces the false positive rate for a modest increase in model complexity, with the added benefit that deriving more accurate shape models also increases the true detection rate (Section 4.5). 

\subsection{Sensitivity to errors in multiple assumed parameters}

While so far we restricted our investigations of the sensitivity of the test to errors affecting a single specific model parameter, when applying the test to real data, almost all model parameters have appreciable error bars. Due to the large number of free parameters in the photometric model, we did not fully characterize the error rates for every possible permutation of parameters. Instead, we explored the effects of perturbing two parameters at a time over a restricted set of values on the true detection and false positive rate. We tested perturbations on all possible pairwise combinations of errors in assumed rotational period, pole orientation, shape model, and phase function, focusing on areas of parameter space (identified in the previous tests) where detection and false positive rates vary most dramatically. Here, we discuss the results of this suite of tests as a whole by examining a two typical examples of models perturbed in two parameters simultaneously. 

\begin{figure}[h!]
    \centering
    \includegraphics[width=\hsize]{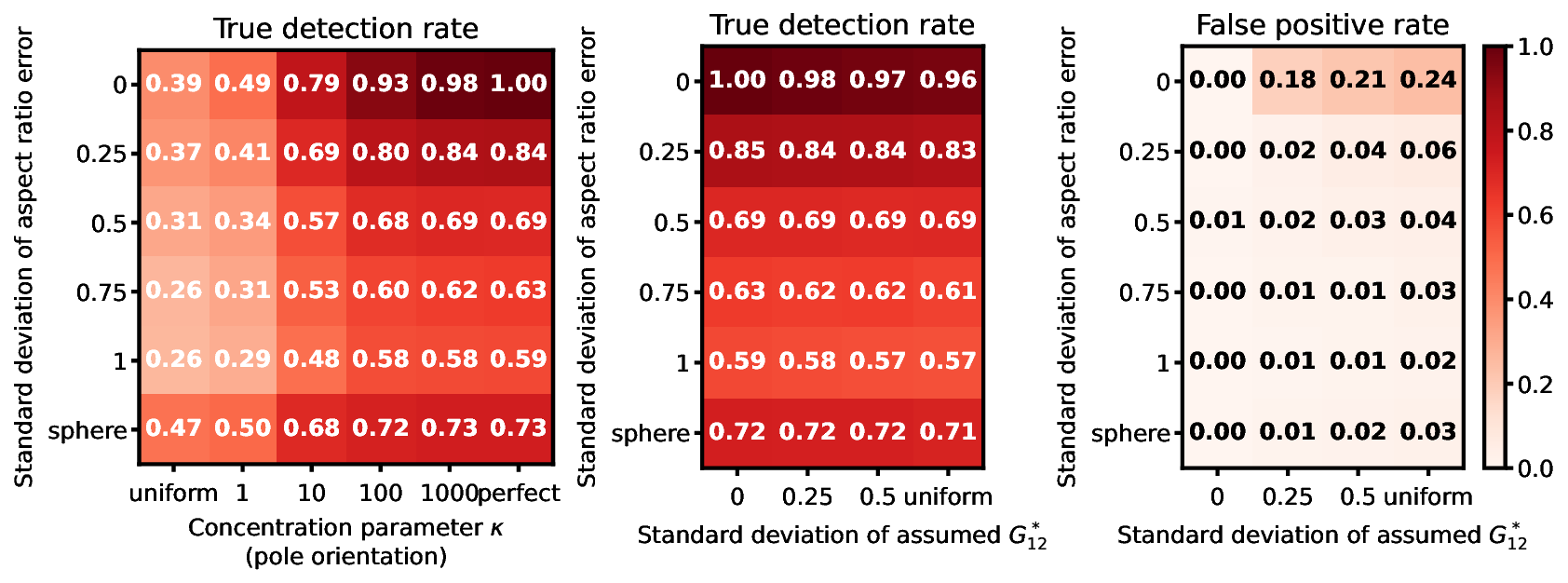}
    \caption{True detection and false positive rates for models simultaneously perturbed in two parameters. In the first panel, the true detection rate is given as a function of error in the assumed pole orientation (parameterized by the concentration parameter $\kappa$) and the standard deviation of the assumed axis ratio error. In the middle panel, the true detection rate is given as a function of the standard deviation of the assumed error in $G_{12}^*$ and the standard deviation of the assumed axis ratio error, as the two parameters were perturbed simultaneously. In the last panel, the false positive rate is given as a function of the standard deviation of the assumed error in $G_{12}^*$ and the standard deviation of the assumed axis ratio error.}
    \label{fig:multi_param}
\end{figure}

Typical examples of the effect of independently varying the error in the assumed values of two model parameters (in this case, shape and pole orientation and shape and $G_{12}^*$) on the detection rate are shown in the first two panels of Figure \ref{fig:multi_param}. Conservatively, the detection rate can be reasonably approximated by multiplying together the expected detection rates in the single-variable perturbation case. For parameters with extremely large errors, this method slightly underestimates the detection rate, reflecting the fact that very large and bright spots that are readily detectable at high errors in one assumed parameter are typically also readily detectable at high errors in the second assumed parameter. 

A typical example of the effect of independently varying the error in the assumed values of $G_{12}^*$ and another model parameter (in this case, pole orientation) on the false positive rate is shown in the final panel of Figure \ref{fig:multi_param}. As in the single variable perturbation test, the false positive rate depends strongly on the error in assumed $G_{12}^*$ value. The false positive rate drops steeply as a function of increasing error in the assumed pole orientation (decreasing $\kappa$). In fact, the ratio of false positives to true detections also decreases as the error in assumed pole orientation increases, indicating that true detections of heterogeneity are more robust to errors in assumed model parameters than false positives. The same is also true for the  true detection and false positive rates under increasing errors in assumed period and pole orientation. Since $G_{12}^*$ is the only perturbed variable with a false positive rate that varies based on the error in its assumed value in the model, all combinations of perturbed model variables not involving perturbations to $G_{12}^*$ showed false positive rates of $<1\%$, no matter the magnitude of the error assumed for either perturbed variable.

\subsection{Practical considerations for applying the test for heterogeneity}

Through simulated datasets, we explored the application and limitations of a statistical test for detecting asteroids with spatially heterogeneous surface colors. We demonstrated that this test can be applied across many populations of small Solar System bodies given the ability to obtain  precise photometric measurements and develop a sufficiently accurate photometric model of an object. A summary of the estimated accuracy in model parameters required to obtain a $>90\%$ detection rate (or, in the case of $G^*_{12}$, a false positive rate of $<10\%$) for each model parameter is summarized in Table \ref{tab:precision}. The accuracy requirements are an average across all simulated spots. In general, larger and higher contrast spots are more easily detectable, while smaller, lower contrast spots require additional accuracy. These accuracy estimations also assume all other model parameters have been correctly estimated. Detection rates for models with errors in multiple parameters can be conservatively estimated by multiplying the detection rates in the single parameter perturbation cases for each parameter.

\begin{table}[]
    \centering
    \caption{Estimated accuracy required for various model parameters.}
    \begin{tabular}{c|c}
         & Maximum error \\
         Parameter & (>90\% detection rate) \\
        \hline 

        Total observations & 85 $\pm$ 30 \\
        Gaussian noise & 0.07 $\pm$ 0.01 (mag) \\
        Rotational period & $3.1 \pm 0. 9 \times 10^{-5} $ \\
        Pole orientation & $7 \pm 2^{\circ}$\\
        Axis ratios & $0.17 \pm 0.04$\\
        \hline
        \hline
         & Maximum error \\
         & (<10\% false positive rate) \\
         \hline
         Phase parameters ($G_{12}^*$)& $0.12\pm0.01$
    \end{tabular}
    \label{tab:precision}
\end{table}

While the test can be flexibly applied to many different small body populations and datasets, it has a few important limitations. First, brightness measurements in each band should be independent. This requirement is met by using non-contemporaneous measurements taken in different photometric bands. For contemporaneous photometric measurements, parameters such as color indices can be computed (or estimated) and the correlation of color correlation with viewing geometry can be assessed directly, as in \cite{Lacerda-2009}, for example. Measurements in different photometric bands should also be drawn from a common distribution of illumination and viewing geometries. Datasets from different time periods (and potentially different surveys) can be combined as long as the coverage of each survey is restricted to those illumination and viewing geometries common to both. Finally, while dense-in-time observations will be of great value for constraining model parameters such as rotational period, we advise against including dense-in-time observations within the datasets evaluated by the statistical test. Introducing densely covered light curves into the datasets considered by the statistical test is prone to increasing the false positive rate when there are errors in the assumed model parameters. The large number of observations provided by dense light curves provide outsized statistical weight to small regions of the light curve that may be poorly modeled and/or otherwise sparsely sampled in other wavelength bands, very similar to the false positives that arise when one band is significantly less well sampled than another.

\section{Conclusions}

We described a computationally straightforward statistical test for detecting color heterogeneity on asteroids using sparse-in-time multiband photometric survey data based on the principle of detecting deviations in light curve shapes as a function of viewing direction in different bandpasses. Applying this test required the ability to obtain $>100$ sufficiently precise photometric measurements of an asteroid under a variety of illumination and viewing conditions. The limiting magnitude of asteroids to which the test may be applied depends on the specific survey strategy and sensitivity of the instruments involved and is mainly restricted by the availability of a sufficient number of observations with small enough magnitude errors in each band and the ability to derive sufficiently precise estimates of the asteroid's basic physical characteristics, including orbit, rotational period, pole orientation, shape, and phase function. Such conditions are met (or are likely to be met in the near future) for a large number of MBAs observed by current telescopic surveys. Current analysis techniques, such as light curve inversion \citep{Kaasalainen-Torppa2001, Kaasalainen-2001} and shape-aware phase function modeling \citep{Carry-2024,Oszkiewicz-2025}, are presently able to obtain the basic characteristics of thousands of asteroids to a sufficient accuracy. Existing survey data pipelines for ZTF and LSST data compute many of these quantities (orbital elements, phase function parameters, pole orientation, and ellipsoidal shape approximations) automatically from sparse data alone as part of their data brokerage services \citep{fink-broker}. These estimates can be used as initial parameter estimates for more sophisticated photometric models, and when available, they can be supplemented by complementary datasets, such as dense light curves, to refine their accuracy. Of all the parameters required to produce a suitable photometric model, achieving the required precision in rotational period is by far the strictest limiting condition determining whether or not our test for color heterogeneity is likely to detect the color variability of a given asteroid. The detection rates of color variations with a strong longitudinal dependence (i.e., color variations in the equatorial region) are most strongly affected by errors in assumed rotational period. The false positive rate is most strongly affected by errors in assumed phase function parameters. Here, recent advancements in asteroid phase function modeling, especially modeling that simultaneously optimizes both global shape models and phase function, are particularly promising for reducing the expected rate of false positives. 

Given the large number of MBAs monitored by large-scale multiband sparse photometric surveys, the widespread application of this test to survey data is likely to yield a statistically robust estimation of the prevalence of regional-scale surface color variability among MBAs. Asteroids identified by the test as strong candidates for displaying color heterogeneity can be further studied by more in depth, targeted follow-up observations. These follow-up observations could include dense color photometry and time-resolved spectroscopy to more precisely constrain the regional distribution of color variations and determine its compositional cause. On the longer term, understanding surface color variation in the asteroid population as a whole will yield insights into near surface geological processes, including space weathering, cratering, and volatile retention, on small Solar System bodies.

\begin{acknowledgements}
      The authors would like to thank the anonymous reviewer for their constructive feedback and comments, which greatly improved the quality of the paper. The authors acknowledge financial support from the VolkswagenStiftung for this research. 
\end{acknowledgements}

\bibliographystyle{aa}
\bibliography{bib}

\begin{appendix}

\section{Details of the photometric model}
The magnitude $m_{B}(t)$ of an arbitrary asteroid at time $t$ in bandpass $B$ was computed via the equation
\begin{equation}
    m_{B}(t)= H_{0, B} + f_{B}(\alpha(t)) + 5 \log (r(t)\delta(t)) -2.5 \log \frac{F_{B}(t)}{F_{0, B}}.
\end{equation}

The initial two terms are related to the standard $H, G$ magnitudes used to parameterize the brightness of small Solar System bodies \citep{Bowell-1989}. Using a similar approach as the restricted definition given in \cite{Carry-2024} to account for the effect of the changing cross-section of a nonspherical body, $H_{0, B}$ is the magnitude of the asteroid as seen at 1 AU, 0$^\circ$ phase angle, with both the subsolar and sub-observer points located at $0^\circ, 0^\circ$ longitude and latitude within the coordinate system of the asteroid. The $f_{B}(\alpha(t))$ term accounts for the dependence of brightness on phase angle $\alpha(t)$ due to the scattering and roughness properties of the surface as in \cite{Kaasalainen-2001}. This function is known to be wavelength dependent, as evidenced by the observation of phase reddening among asteroids \citep{Mahlke-2021,Alvarez-Candal-2024, Wilawer-2024}. Therefore, for each bandpass of interest, a corresponding phase function and absolute magnitude is required. The specific phase function adopted here can in principle be computed according to one of many common scattering models ($HG$, $HG_1G_2, HG_{12}$, $HG^*_{12}$ or the linear-exponential model, see \citet{Kaasalainen-lin-exp-2001, Muinonen-2010, Penttila-2016}). The $5 \log (r(t)\delta(t))$ term corresponds to the standard correction from reduced magnitude to apparent magnitude for Solar System objects, with $r(t)$ and $\delta(t)$ representing the heliocentric and observer-centric distances in AU as computed via ephemeris. 

Finally the $-2.5 \log \frac{F_{B}(t)}{F_{0, B}}$ term accounts for the difference in brightness due to the changing cross-sectional area and illumination state of the asteroid. Analogously to $H_{0, B}$, $F_{0, B}$ is the computed flux from the asteroid as seen at 1 AU, 0$^\circ$ phase angle, with both the subsolar and sub-observer points located at $0^\circ, 0^\circ$ longitude and latitude within the coordinate system of the asteroid. $F_B(t)$ is the flux from the asteroid that would be observed given its actual orientation relative to the Sun and observer at time $t$, effectively adjusting $H_{0,B}$ to the viewing geometry-specific $H$ magnitude at a given $t$. Computing this quantity requires a model of the asteroid's wavelength-invariant shape and a wavelength-dependent reflectance map. We assume that an asteroid's shape can be represented as a convex triangular mesh, with each triangular facet $T$ assigned a reflectance $R_{T, B}$. For a uniformly colored asteroid, $R_{T,B}$ will be constant in each band. The convexity assumption significantly simplifies flux calculations by ensuring self-shadowing and self-illumination can be ignored. Further, convex approximations of asteroid shapes, even for asteroids with significant non-convexities, are adequate to provide sufficiently accurate approximations of light curves in the context of photometric modeling \citep{Kaasalainen-Torppa2001}. Under these assumptions, we then computed the total flux from the shape and reflectance map over all triangular facets using the equation

\begin{equation}
    F_{B}(t) = \sum_{T}{A_{T}R_{T, B}  S(\mu_T(t), \mu_{0, T}(t)) },
\end{equation}

where the area of each facet is given by $A_{T}$ and the disk function $S(\mu_T(t), \mu_{0, T}(t))$. As in \cite{Kaasalainen-2001}, $\mu_T(t)$ and $\mu_{0, T}(t)$ are the dot product of the surface normal of each facet and the unit vector (in the rotating frame of the asteroid) pointing in the direction of the observer ($\hat{E}(t)$) and Sun ($\hat{E_0}(t)$), respectively. We employed a combined Lambertian and Lommel-Seeliger scattering law as in the equation

\begin{equation}
    S(\mu_T(t), \mu_{0, T}(t)) = \begin{cases}
    \mu_T \mu_{0,T}\left(\frac{1}{\mu_T + \mu_{0, T}} + 0.1 \right) & \text{if $\mu_{0, T} > 0 $ and $ \mu_T > 0$} \\
    0 & \text{if $\mu_{0, T} \leq 0 $ or $ \mu_T \leq 0$}\\ 
    \end{cases}.
\end{equation}

The piecewise function arises from the observation that a negative value of $\mu_{0, T}$ ($\mu_{T}$) implies, under the convexity assumption, that the Sun (observer) is behind the facet, and that facet is unlit (invisible to the observer), thus contributing zero flux. Given an asteroid with a known rotational period and rotational pole orientation in ecliptic coordinates, the unit vectors $\hat{E_0}$ and $\hat{E}$ in the rotational frame of the asteroid can be computed straightforwardly from ephemeris data by reversing the series of rotations described in Equation 1 of \cite{Durech-2010}. 

The model was implemented in python, using the \textit{sbpy} \citep{sbpy-2019} and \textit{phunk} packages for phase function calculations and the \textit{kete} package \citep{kete-2025} for ephemeris calculations. The packages \textit{scipy}, \textit{astropy}, \textit{pandas}, \textit{scikit-learn}, \textit{matplotlib}, and \textit{numpy} were used in the technical implementation of the model and production of figures in the text.

\FloatBarrier
\clearpage

\end{appendix}
\end{document}